\documentstyle[aps,preprint,psfig]{revtex}

\def\be{\begin{eqnarray}}
\def\ee{\end{eqnarray}}

\def\ket#1{| #1 \rangle}
\def\bra#1{\langle #1 |}
\def\r{\rangle}
\def\l{\langle}
\def\ad{a^{\dagger}}
\def\bd{b^{\dagger}}

\begin{document}

%\twocolumn

\title{Polarization in Quantum Computations}
\author{P\"aivi T\"orm\"a and Stig Stenholm$^*$\\
%EndAName
Research Institute for Theoretical Physics,
P.O.Box 9, 
%(Siltavuorenpenger 20C)
FIN-00014 UNIVERSITY OF HELSINKI,
Finland}
\date{January 5, 1996}
\maketitle

\begin{abstract}
We propose a realization of quantum computing using polarized photons.
The information is coded in two polarization directions of the photons
and two-qubit operations are done using conditional Faraday effect. 
We investigate the performance of the system as a computing device.
\end{abstract}
\vspace{0.3cm}
PACS numbers: 03.65.Bz, 89.80.+h, 32.80.-t, 42.50.-p
\vspace{0.4cm}

\pagebreak

After the early discussion of quantum computing
\cite{Feynman82,Deutsch85}, the field has attracted much attention
because Shor \cite{Shor94} has shown that the famous factorization
problem can, in principle, be speeded up considerably by quantum data
manipulation techniques. The recent work on quantum computations has
been reviewed by Bennett in Ref.\cite{Bennett95}. Many realizations
have been suggested, at present the most promising seem to be ions
trapped electrodynamically \cite{Cirac95} or in a cavity
\cite{Pellizzari95}. 

In the recent work \cite{Stenholm95}, one of us considered the possible
use of photon polarization states to carry quantum information. The
advantage is that they provide a natural two-state basis with no
additional Hilbert space components, such as the vacuum state, 
that may constitute losses of the
coding. The single photon coding allows an easy detection,
in contrast the vacuum state is hard to distinguish from a
failed detection.
The photon coding also allows long dephasing times and the
possibility to transfer the information from one device to another
through fibers. The purpose of this paper is to investigate
how realistic this suggestion is by numerical integration of a
semirealistic situation.

The elliptically polarized photon state $(\alpha_+ \ad_+ + \alpha_-
\ad_-) \ket 0$ can be manipulated by the Faraday 
effect induced by the presence of a second photon $b_{\pm}^{\dagger}$. 
These are supposed to
selectively transfer population from the ground state $\ket 0$ in Fig.1
to the levels $\ket{\pm 1}$. Because each photon $a_{\pm}^{\dagger}$
sees only one transition $\ket{\pm 1} \rightarrow \ket 2$ it becomes
modified by the population transferred by the photons
$b_{\pm}^{\dagger}$. If we keep the transition $\ket{\pm 1} \rightarrow
\ket 2$ off resonance, the atom acts as a dielectric only and hence the
relative phases of $a_{\pm}^{\dagger}$ are modified; this is a turning
of the axis of the state $(\alpha_+ \ad_+ +
\alpha_- \ad_-) \ket 0$. It was shown in
\cite{Stenholm95} that this allows the gated application of an arbitrary
unitary transformation.

In this paper we are looking at two different cases. Case I corresponds
to Fig.1 when $\Delta_1=0$. 
Here both transitions $\ket{0} \rightarrow \ket{\pm 1}$ are in
resonance and both photons $a_{\pm}^{\dagger}$ experince a modified
phase. The situation is symmetric: if $b_+^{\dagger}$ is present alone
we achieve a phase shift exactly opposite in sign to that caused by the
presence of $b_-^{\dagger}$ only.
In Case II we detune one of the transitions, $\ket{0} \rightarrow
\ket{+1}$ say; see Fig.1. Then only the presence of the resonant photon
$b_+^{\dagger}$ affects the phase of the $a^{\dagger}$-photons. This
corresponds to the gate where the presence of the state
$b_-^{\dagger}\ket{0}$ does nothing. Most gates discussed earlier in the
literature are of this type.

The four-level system shown in Fig.1 is described by the Hamiltonian 
\be
H&=&\Omega _2( a_{+}^{\dagger }a_{+}+a_{-}^{\dagger }a_{-})
+\Omega
_1( b_{+}^{\dagger }b_{+}+b_{-}^{\dagger }b_{-}) \nonumber \\ 
&&+\omega _2\ket 2 \bra 2 +\omega_{1+}\ket{+1}
\bra{+1}+\omega_{1-}\ket{-1}\bra{-1}  \nonumber  \\
&& +\omega _0 \ket 0\bra 0 +
\lambda _1\left( b_{-}\ket{+1}\bra 0 +b_{+}\ket{-1}
\bra 0 +h.c.\right)  \nonumber \\ 
&& + \lambda _2\left( a_{-}\ket 2\bra{-1}+a_{+}\ket 2
\bra{+1} +h.c.\right) .  \label{hamiltonian}
\ee
In Case I we assume that the states $\ket{\pm 1}$ are degenerate and
that the transitions $\ket{0} \rightarrow \ket{\pm 1}$ are at
resonance, $\Delta_1^I = \omega_{1\pm} - \omega_0 - \Omega_1 = 0$.
The transitions $\ket{\pm 1} \rightarrow \ket 2$ are assumed detuned, i.e.
$\Delta_2^I = \omega_2 - \omega_0 - \Omega_1 - \Omega_2 = \omega_2 
- \omega_{1\pm}- \Omega_2$ is nonzero.
In Case II we lift the degeneracy of levels $\ket{\pm 1}$ by
setting $\omega_{1+}\ne \omega_{1-}$. Then the transition $\ket 0
\rightarrow \ket{-1}$ is taken at resonance
$\omega_{1-}-\omega_{0}=\Omega_1$ but the detunings
$\Delta_1^{II} = \omega_{1+}-\omega_0 - \Omega_1$ and 
$\Delta_2^{II} = \omega_2 - \omega_0 - \Omega_1 -\Omega_2$
are nonzero. The transition $\ket{+1} \rightarrow \ket 2$ is detuned by
${\Delta'}_2 = \omega_2 - \omega_{1+} - \Omega_2 
= \Delta^{II}_2 - \Delta^{II}_1$;
this is assumed well off resonance too.

The initial state is taken to be the disentangled form
\be
\ket{\Psi_{in}}=(\alpha_+ a_+^{\dagger}+\alpha_- a_-^{\dagger})
(\beta_+ b_+^{\dagger} + \beta_- b_-^{\dagger}) \ket 0  , \label{statevec}
\ee
where $\ket 0$ denotes the vacuum of the fields. The coefficients are
in general complex numbers normalized to unity. 
We propagate the state vector (\ref{statevec}) to the time $t$ with
the Hamiltonian (\ref{hamiltonian})  and write the final state as
\be
\ket{\Psi_{out}} = e^{-i H t} \ket{\Psi_{in}} = \sum_{i=1}^9 C_i \ket
i  ,  \label{final}
\ee
where we have numbered the basis states according to the set
\be
&&\{ \ket 1 ,\ket 2 ,\ket 3 , \ldots , \ket 9 \} = 
\{ \ket 2 , a_+^{\dagger}\ket{+1}, a_+^{\dagger}\ket{-1},
a_-^{\dagger}\ket{+1}, \nonumber \\ && a_-^{\dagger}\ket{-1}, 
a_+^{\dagger} b_+^{\dagger}\ket 0 , a_+^{\dagger} b_-^{\dagger} \ket
0 , a_-^{\dagger} b_+^{\dagger} \ket 0 , a_-^{\dagger}
b_-^{\dagger}\ket 0 \}  .
\ee
Initially the coefficients $\{C_6,C_7,C_8,C_9\}$ are prepared. Of
these, the Hamiltonian couples in Case I $C_6$ to $C_3$ and $C_9$ to $C_4$
only; in Case II $C_6$ to $C_3$ only. 
In these subspaces the system can be solved exactly, and
performing a rotating wave approximation with respect to the
frequency $\omega_0+\Omega_1+\Omega_2$ we obtain in Case I
\be
C_9(t) &=& \cos (\lambda_1 t) C_9(0) + i \sin (\lambda_1 t) C_4(0) \\
C_6(t) &=& \cos (\lambda_1 t) C_6(0) + i \sin (\lambda_1 t) C_3(0) ; 
\label{C6}
\ee
in Case II only (\ref{C6}) is valid.
Choosing the interaction time such that $\lambda_1 t=\pi$, we find
that the probabilities are restored in these subspaces.
We are now left in Case I 
with a 5 dimensional and in Case II a 7 dimensional 
subspace to consider numerically.

After the interaction, the state (\ref{final}) is available for
measurements. In the ideal situation, the initial photons would have
been restored to the radiation field. This is desired because the
information resides in these photons, and they should be available
for subsequent computational operations. We can ensure that they have
been returned by observing that the atom is back in its ground state
$\ket 0$ by projecting the final state on this. After the
interaction, the atom is available for inspection; a measurement on
its state does no longer affect the outcome of the process. We write
this state after an observation, $\ket{\Psi_0} = \ket 0 \l 0 |
\Psi_{out}\r$, as
\be
&&\ket{\Psi_0} = \left( \frac{C_{++}}{\alpha_+ \beta_+}
e^{i\varphi_{++}} \alpha_+ a_+^{\dagger} + \frac{C_{-+}}{\alpha_-
\beta_+} e^{i\varphi_{-+}}\alpha_- a_-^{\dagger} \right) \beta_+ \bd_+ \ket
0 \nonumber \\
&&+ \left( \frac{C_{+-}}{\alpha_+ \beta_-} e^{i\varphi_{+-}} \alpha_+
\ad_+ + \frac{C_{--}}{\alpha_- \beta_-} e^{i\varphi_{--}}\alpha_- \ad_-
\right) \beta_- \bd_- \ket 0  . \label{final2}
\ee
We have written the amplitudes and phases of the new coefficients as $C_{ij}
e^{i\varphi_{ij}}$ ($i,j\in \{-,+\}$).

A measure of the efficiency of the process is the probability 
$P_0 = |\l 0 | \Psi_{out}\r |^2$.
A small value of $P_0$ makes the process inefficient, but once the
state $\ket 0$ has been observed on the atom, the expressions in the
brackets of (\ref{final2}) give the effect on the state $(\alpha_+
a_+^{\dagger} + \alpha_- a_-^{\dagger}) \ket 0$ conditioned on the
presence of the photons $b_{\pm}^{\dagger}$ on the lower transitions.
These expressions contain the effect of the gating action of the
system. In all cases investigated in this paper, however, $P_0$ has
been found to deviate from unity by less than 1\%. The process is efficient
as given.

If the coefficients $\eta_{ij} = C_{ij}/|\alpha_i \beta_j|$
in (\ref{final2}) 
are close to unity, the interaction only adds the phases $\varphi_{ij}$;
the polarization of the $\ad$-field has been changed by the
interaction. If we define the initial phases $\varphi_{\pm}^a =
arg(\alpha_{\pm})$ and $\varphi_{\pm}^b = arg(\beta_{\pm})$, we denote the
phase changes by
\be
\overline{\varphi }_{\pm }&=& ( \varphi _{+\pm }+\varphi
_{-\pm
}-\varphi _{+}^a-\varphi _{-}^a)/2 -\varphi _{\pm }^b 
\label{phase2} \\ 
\Delta \varphi _{\pm }&=& ( \varphi _{+\pm }-\varphi _{-\pm
}-\varphi _{+}^a+\varphi _{-}^a)/2 . \label{phase1}
\ee

We now write the final state (\ref{final2}) in the form
\be
&&\ket{\Psi _0} = \left\{ e^{i\overline{\varphi }_{+}}\left(
\eta
_{++}\,e^{i\Delta \varphi _{+}}\alpha _{+}\,a_{+}^{\dagger }+\,\eta
_{-+}e^{-i\Delta \varphi _{+}}\alpha _{-}\,a_{-}^{\dagger }\right) \beta
_{+}b_{+}^{\dagger }\right. \nonumber \\ 
&  & \left. +e^{i\overline{\varphi }_{-}}\left( \eta _{+-}\,e^{i\Delta
\varphi _{-}}\,\alpha _{+}a_{+}^{\dagger }+\eta _{--}e^{-i\Delta \varphi
_{-}}\alpha _{-}\,a_{-}^{\dagger }\right) \beta _{-}b_{-}^{\dagger
}\right\}
\ket 0 . \nonumber
\ee
When we choose the initial coefficients $\alpha_{\pm},\beta_{\pm}$
real, the phases (\ref{phase2})--(\ref{phase1}) 
simplify; at the end of this paper we are going to
discuss the influence of the phase on the gating performance.

In Case I, the symmetry requires that $\overline{\varphi}_+ 
=\overline{\varphi}_-$
and $\Delta \varphi_+ = -\Delta \varphi_- \equiv \Delta \varphi$. In Case II,
we assert that $\varphi_{+-}\sim \varphi_{--}\simeq0$, which implies
$\overline{\varphi}_- \simeq 0$ and $\Delta \varphi_- \simeq 0$. We may consider 
the 4-dimensional subspace 
$\{a_{-}^{\dagger }b_{-}^{\dagger }\ket 0 ,
a_{+}^{\dagger }b_{-}^{\dagger }\ket 0 ,
a_{-}^{\dagger }b_{+}^{\dagger }\ket 0 ,
a_{+}^{\dagger }b_{+}^{\dagger }\ket 0 \}$.
Assuming now that all coefficients $\eta_{ij}$ are unity, we obtain
in the symmetric case the ideal transformation $U_I\sim 
e^{i\overline{\varphi }} Diag\{e^{i\Delta \varphi }, e^{-i\Delta \varphi
}, e^{-i\Delta \varphi }, e^{i\Delta \varphi }\}$. In the detuned Case II,
we obtain $U_{II}\sim Diag\{1,1,e^{i(\overline{\varphi}_+ +\Delta
\varphi_+)}, e^{i(\overline{\varphi}_+ - \Delta \varphi_+) }\}$.
This is a phase transformation of the bit denoted by $\ad_{\pm}$
induced by the presence of the photon $\bd_+$. 

We are now going to consider the performance qualities of the model
system as a gated bit transformation. The input to the calculation is
the initial state (\ref{statevec}). To begin we choose the
"classical" case when only one of the input states is present.
In the symmetric 
Case I, the choice of state is not important, c.f.
$U_I$, but for the Case II, we need to look at the
states $\ad_- \bd_+\ket{0}$ and $\ad_+ \bd_+\ket{0}$. 
First we choose to discuss the
single input state $\ad_- \bd_+\ket{0}$ with $\alpha_-=\beta_+=1$.

As stated above, the interaction time is chosen such that
$t=\pi/\lambda_1$; in the calculations we choose $\lambda_1=1$. For
large detunings ($\omega _2\rightarrow \infty $) $\eta _{-+}$ approaches
unity, 
but the phase shift $\Delta \varphi $ goes to zero. In Case I, the
numerical
investigations show that we can retain $\eta^2_{-+}>0.9$ if we choose
$\Delta_2^I >5$. For $\Delta_2^I = 5$ we find $\Delta \varphi_+ \simeq
10^{\circ }$. This is achieved with $\lambda _2=1;$
larger phases can be achieved by increasing $\lambda _2,$ but the
restoring
of the population suffers. 
For $\lambda _2 \le 1.5$ we can achieve $\Delta \varphi_+ \ge 
15^{\circ }$ and 
$\eta^2_{-+}>0.75$.
The results can be illustrated
in a graph plotting $\Delta \varphi_+$ as a function
of $\eta_{-+}^2$ with the detuning as a parameter. For the symmetric Case
I, this is done in Fig.2a. As we can see, for $\Delta_2^I > 5$, no
dependence on detuning is seen. The corresponding results for Case II
are shown in Fig.2b. Here the dependence on detuning is much stronger;
however for large values of detuning, $\Delta_1^{II} = 15$ and
$\Delta_2^{II} = 30$, we can reach $\Delta \varphi_+ \ge 43^{\circ }$ with
$\eta_{-+}^2 \ge 0.9$. Thus the operation of this gate is much more
efficient as is to be expected. For larger values of $\Delta_1^{II}$ the
results tend to become independent of the detuning.

We now choose to look at the case $\lambda _2=2.5$ and $\Delta_2 = 30$.
For the Case I this gives $\Delta \varphi_+ \simeq 10^{\circ }$ and
$\eta_{-+}^2 \simeq 0.90$. In Case II it gives $\Delta \varphi_+ \simeq
10^{\circ }$ and $\eta_{-+}^2 \simeq 0.99$. In order to see where the
missing population goes in Case I, we plot the population of the states
$a_{-}^{\dagger }b_{+}^{\dagger }\ket 0$, $a_{-}^{\dagger }\ket
0$, $a_{+}^{\dagger }\ket 0$, 
$a_{+}^{\dagger }b_{-}^{\dagger }\ket 0$ and $\ket 2$ in 
Fig.3. At time $t=\pi$, the population of
$a_{-}^{\dagger }b_{+}^{\dagger }\ket 0$ is restored to 90\% but
the missing population is on the level 
$a_{+}^{\dagger }b_{-}^{\dagger }\ket
0$. This is mediated through the
off-resonant transition $\ket{-1} \rightarrow \ket 2 \rightarrow
\ket{+1}$ which proceeds at the effective Rabi rate $(\lambda_2^2
/\Delta_2^I ) \sim 6.25/30$. With time, this increases the
population of the state $a_{+}^{\dagger }\ket{+1}$ as can be seen
in Fig.3; this increase is modulated at the rate $\lambda_1$ by the
population pulsations on level $a_-^{\dagger }\ket{-1}$. This
effect can be decreased by increasing $\Delta_2 >> \lambda_2^2$.
In Case II, the population of the level $a_{-}^{\dagger
}b_{+}^{\dagger }\ket 0$ is restored to better than 99\% and the
population on states $a_{+}^{\dagger }\ket{+1}$ and
$a_{+}^{\dagger }b_{-}^{\dagger }\ket 0$ remain below $10^{-3}$.

After having described the ''classical'' inputs, where each 2-bit pure
state has been treated separately, we now turn to consider the genuine
quantum situation described by the input state (\ref{statevec}). The
performance of the system acting on this state is, of course, essential
for its usefulness as a quantum computing device.

An input consisting of a pair of two-level systems contains 4 degrees of
freedom: the 4 complex numbers involved loose two parameters to the
over-all phase and two to the normalization conditions. It is still
difficult to display the results of a 4 parameter input space, and hence
we start by considering only real coefficients in Eq.(\ref{statevec}).
The influence of the phases $\varphi_{\pm}^a$, $\varphi_{\pm}^b$ will be
discussed below. 
We are thus left with two real parameters, one for each input bit. We
choose to display our results as functions of 
$\alpha_-^2=1-\alpha_+^2$ for the two cases
\be
\begin{array}{ccc}
\ket{\beta _1}& = & \frac 1{\sqrt{2}}\left( b_{+}^{\dagger
}+b_{-}^{\dagger }\right) \ket 0 \\ 
\ket{\beta _2} & = & \left( \frac{\sqrt{3}}2b_{+}^{\dagger }+\frac
12b_{-}^{\dagger }\right) \ket 0 
\end{array}  .
\ee

We want to introduce a quality factor for the use of a system
like this in computations. The performance is close to ideal, when the
parameter $\eta_{ij}\simeq 1$. However, when either one of the input
parameters $\alpha_i,\beta_j$ becomes close to zero, any minute value
in the corresponding coefficient $C_{ij}$ is likely to cause a large
value $\eta_{ij}$. Thus we want to consider the retention of that
product $\alpha_i\beta_j$ which is the largest. A value close to unity
here signals a good performance. To test this idea we consider the variables
\be
\eta_{-+}^2 \quad 
(\alpha_-^2 \ge 0.5) \quad ; \quad 
\eta_{+-}^2 \quad
(\alpha_-^2 \le 0.5)  .  \label{quality}
\ee
Another measure of the efficiency of the process can be given by the
retention of the ratio between the two components $\bd_{\pm}$ in
Eq.(\ref{final2}). This starts from $|\beta_+ /\beta_-|^2$
and if retained the parameter
\be
R=\left( \frac{\mid C_{++}\mid ^2+\mid C_{-+}\mid ^2}{\mid C_{+-}\mid
^2+\mid C_{--}\mid ^2}\right) \left( \frac{\beta _{-}}{\beta
_{+}}\right)^2 
\ee
should be close to unity. 
The retention parameter $R$ for Case I and the inputs $\ket{\beta_1}$
and $\ket{\beta_2}$ is shown in Fig.4a together with the corresponding
quality factor in Eq.(\ref{quality}). In Fig.4b the same parameters are shown for
the asymmetric Case II. As we can see, the retention parameter $R$ is
at its worst about 70\%; in Case II it is better than 90\%. In Case I,
the quality factor (\ref{quality}) is good to within 90\% and in the asymmetric Case
II to better than 95\%.

Finally we want to return to the question of the influence of the
initial phases. These do affect the outcome, but their influence
seems to be smaller than the influence of the magnitudes. 
We consider the 
achieved phase shifts as functions of the superposition
coefficients $\alpha$ and $\beta$. 
In Fig.5 we plot the phase shifts $\Delta \varphi_{\pm}$ 
against
$\alpha_-^2$ in the asymmetric Case II
shown for $\ket{\beta_1}$ and $\ket{\beta_2}$.
For $\ket{\beta_1}$, we also consider the case when 
the initial phase $\varphi_+^a$ is set to the value $\varphi_+^a=\pi/4$.
The behaviour is close to ideal; in the range $\alpha_-^2 \in (0.1,0.9)$,
nearly ideal
behaviour is observed, $\Delta\varphi_+ \simeq 9.5^{\circ}$ and
$|\Delta\varphi_- | < 0.4^{\circ}$. The effect of the initial phase is
small. In the symmetric Case I, the behaviour was found to be less optimal:
we saw only a small difference
for the two $\beta$-states, but for $\alpha_-^2$ in the range
$(0.1,0.9)$ the phase shift changed from $30^{\circ}$ to
$10^{\circ}$. Thus in Case I, the magnitude of the angle 
remains considerable but it
does depend on the value of $\alpha$. 
We have not carried out a systematic investigation of the 
influence of the phase factors;
the results reported here indicate that they cause no drastic
changes. If needed, their effects can easily be evaluated using
the method presented here. 

As a conclusion, we discuss how well a quantum gate can be realized 
in our model. We choose to look at the Controlled-NOT gate, which
changes the value of the target bit whenever the control bit has the
value one. Based on the considerations above, we conclude that the 
asymmetric Case II is better suited to work as a gate. Its performance
can easily be improved from the results 
presented above by increasing $\Delta_2^{II}$, $\Delta_1^{II}$ and $\lambda_2$
in a suitable way. Here we use the parameters
$\Delta_2^{II}=70$, $\Delta_1^{II}=65$, $\lambda_2=6.85$, $\lambda_1=2$,
and $t=\pi$: this enables us to approximate the transformation $U_{II}$
to the accuracy $10^{-3}$ with a phase shift of $60^{\circ}$.
This has to be applied three times in sequence in order to get a phase
shift of $180^{\circ}$, which is needed for the Controlled-NOT gate. 
After performing suitable transformations
between the circular and linear bases (see \cite{Stenholm95}), we obtain as the
final result the Controlled-NOT transformation $C_N$:
\be
C_{N} = \left[\begin{array}{cccc}
0.995 e^{-i 33^{\circ}} & {\cal O}(10^{-3}) & {\cal O}(10^{-2}) &
{\cal O}(10^{-2}) \\
{\cal O}(10^{-3}) & 0.995 e^{-i 33^{\circ}} & {\cal O}(10^{-2}) &
{\cal O}(10^{-2}) \\
{\cal O}(10^{-2}) &  {\cal O}(10^{-2}) & {\cal O}(10^{-3}) &
-0.997 \\
{\cal O}(10^{-2}) & {\cal O}(10^{-2}) & -0.997 & {\cal O}(10^{-3})
\end{array} \right]  .  \nonumber
\ee
The overall phases $e^{-i33^{\circ}}$ and $-1$ are irrelevant. We see that
the Controlled-NOT gate can be realized in this case to the accuracy $10^{-2}$.

The present scheme has been found to perform reasonably well as a
computing device. It is naturally not good enough to be an element of a
computer network of realistic size, but there seems to be no suggestion
in the literature which satisfies this criterion. The performance of our
scheme can be improved by sequantial application of the $\bd$- and
$\ad$-photons, with final restoration of the $\bd$-state by a third
pulse. Such a scheme seems to require perfectly controlled pulses, which
we regard as even more unrealistic than the model we have investigated.
To implement our method in a multi-step computation we assume all
initial information to be coded in a set of field modes residing
uncoupled in the same cavity. During their coherence time, we shoot
through the cavity volume a sequence of suitably chosen atoms which
couple the photon pairs, i.e. perform the two-qubit operations. To
affect all possible unitary transformations, the cavity has to be rather
complicated, containing a suitable arrangement of $\lambda$-plates to
give access to all desired polarization states. Also the atoms have to be
able to couple just the desired modes at each stage of the calculation.
This and the restrictions imposed by loss rates and decoherence times
pose extremely strict limitations on the computations possible. If
several cavities are necessary, the dissipative effects on photons
transferred between them raise further problems. However, such
difficulties seem to afflict other schemes suggested too. Which one can
be optimized the most remains an experimental challenge.

\pagebreak

\begin{figure}[htb]

\centerline{
\psfig{figure=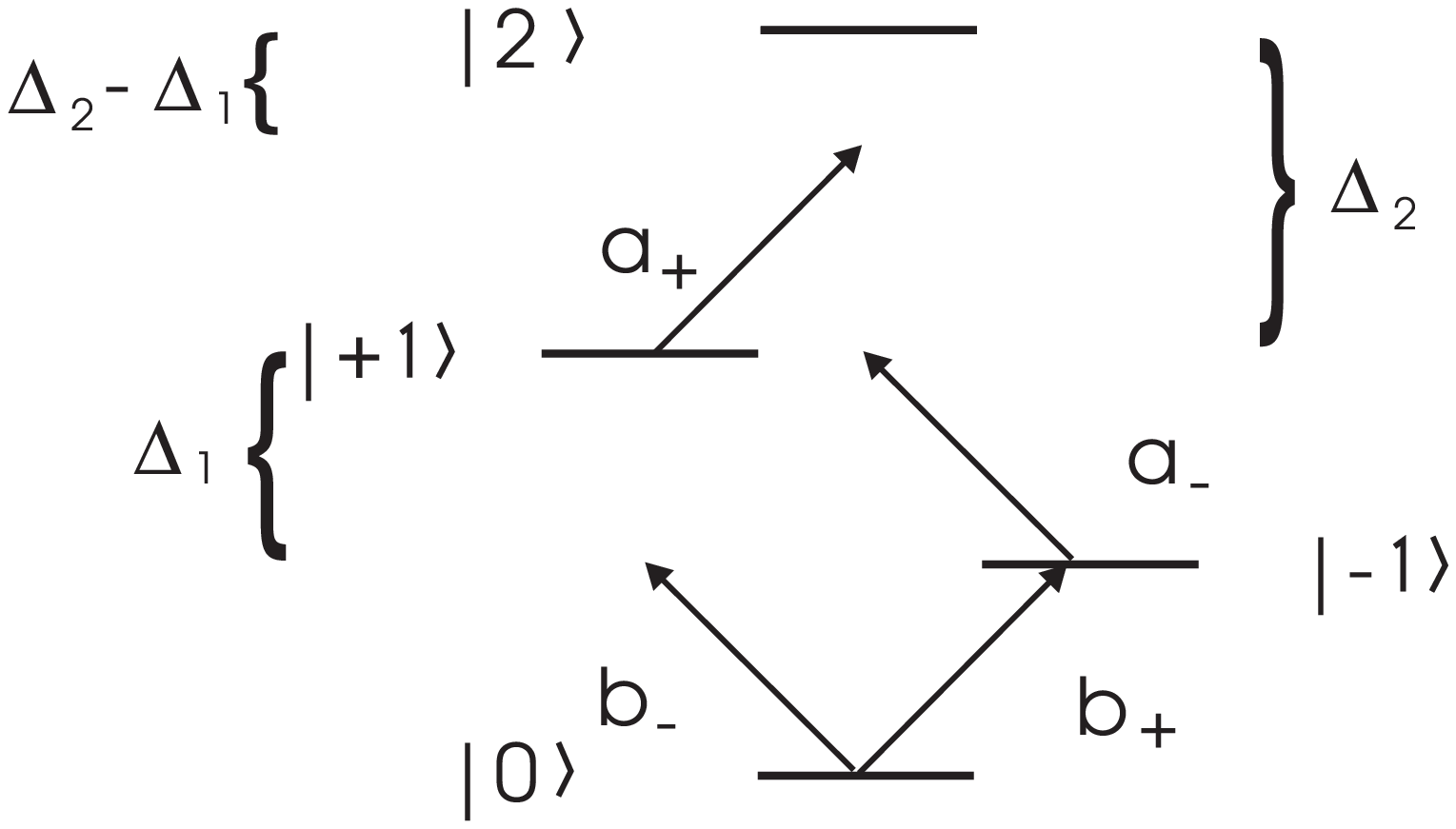}
}
\caption{The 4-level system used in the gated Faraday effect}

\pagebreak

\centerline{
\psfig{figure=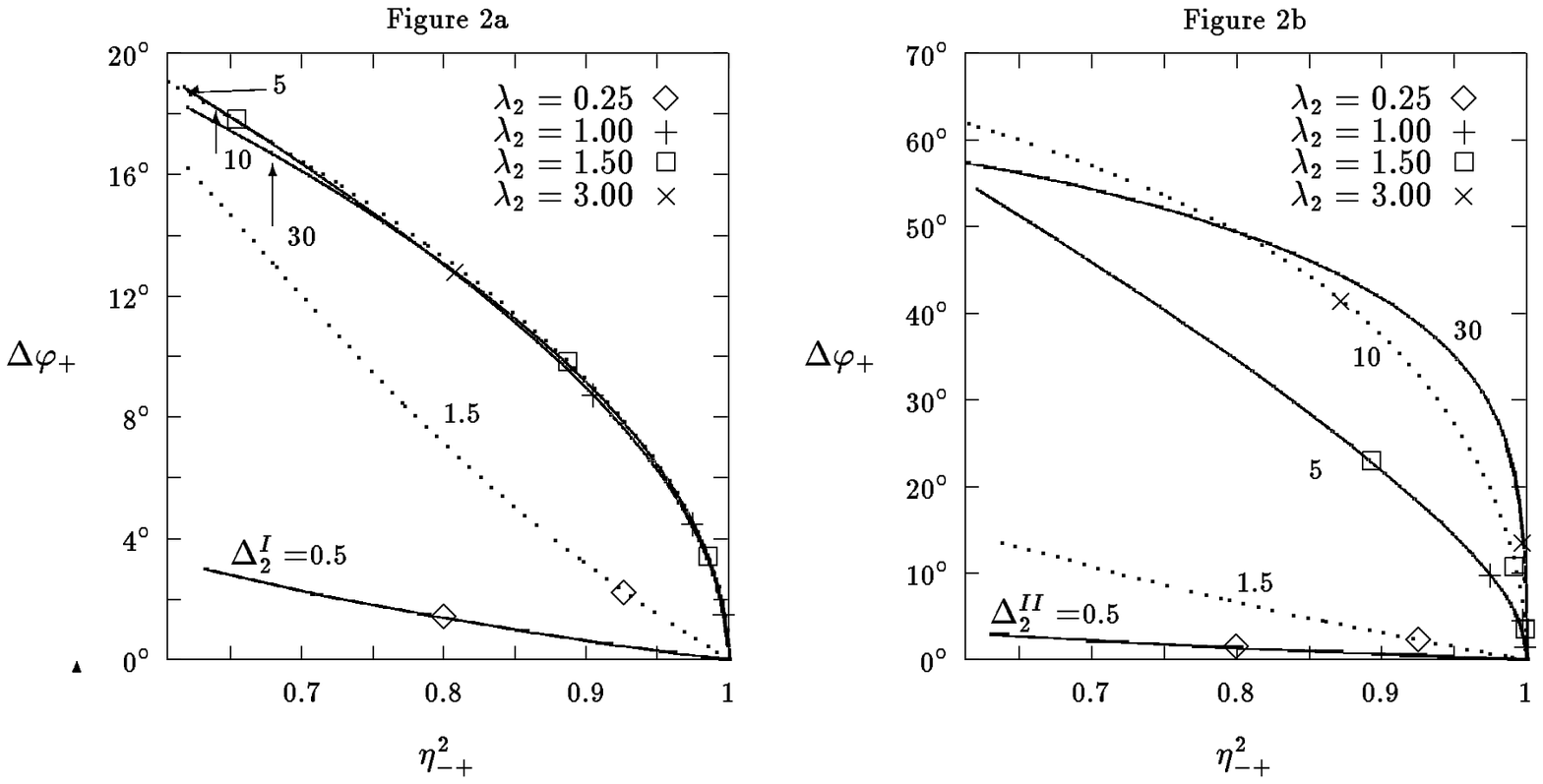}
}
\caption{The phase shift $\Delta \varphi_+$ as a function of 
$\eta_{-+}^2$ for several values of the detuning $\Delta_2^I$, $\Delta_2^{II}$;
some values of $\lambda_2$ used are marked.
In all figures Fig.\#a corresponds to Case I, Fig.\#b to Case II where
$\Delta_1^{II}=\Delta_2^{II}/2$.}

\pagebreak

\centerline{
\psfig{figure=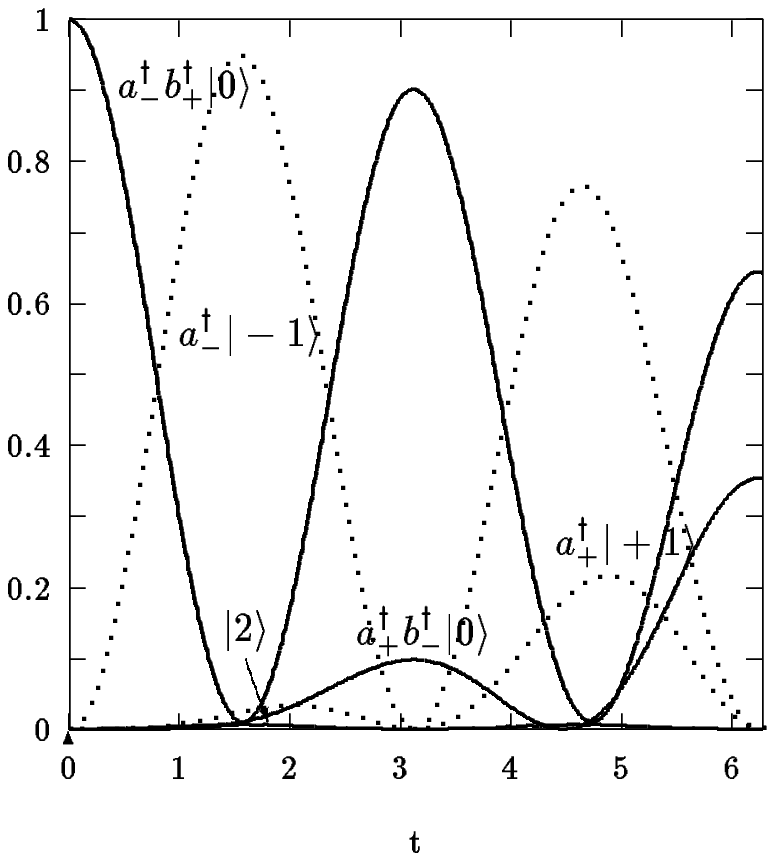}
}
\caption{The populations of the basis states as functions of time (Case I)}

\pagebreak

\centerline{
\psfig{figure=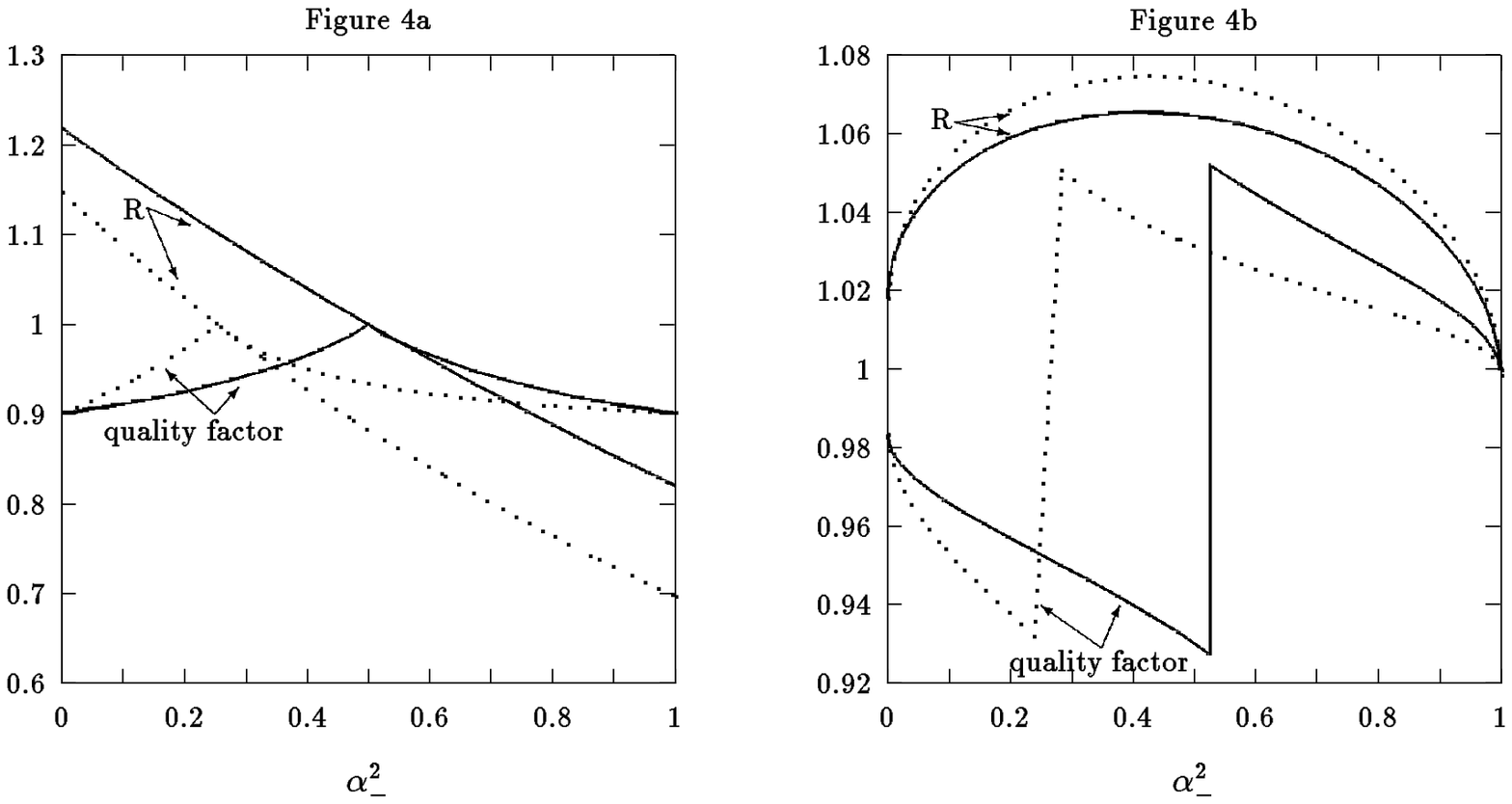}
}
\caption{The retention $R$ and the quality factor (20)
as functions of $\alpha_-^2$, for $\ket{\beta_1}$ (solid lines) and
$\ket{\beta_2}$ (dotted lines)}

\pagebreak

\centerline{
\psfig{figure=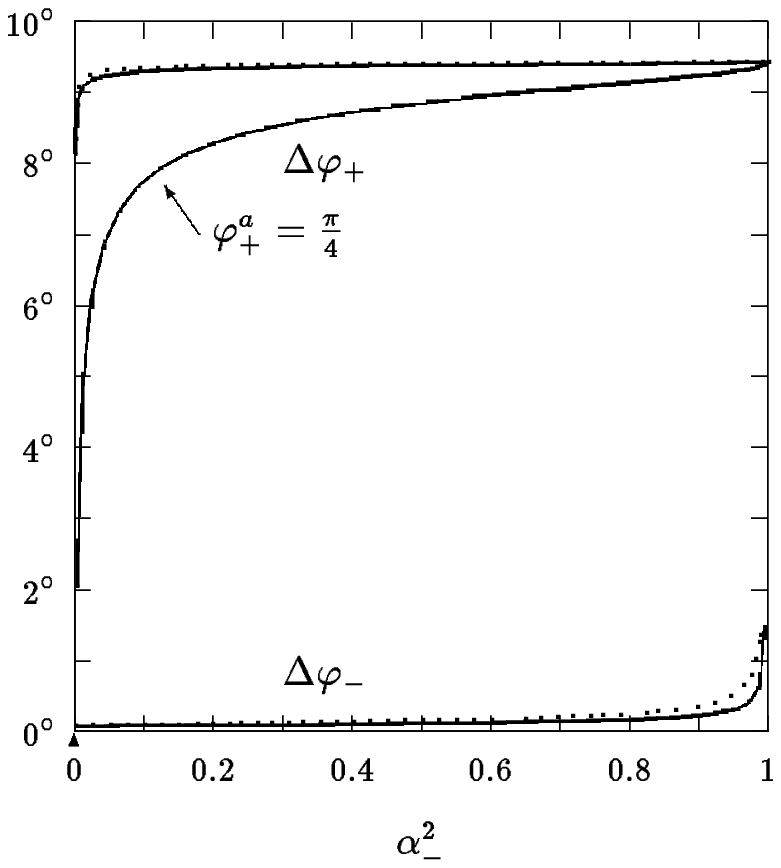}
}
\caption{The phase shifts $\Delta\varphi_{\pm}$ as functions of
$\alpha_-^2$, for $\ket{\beta_1}$ (solid lines) and  
$\ket{\beta_2}$ (dotted lines). The shift $\Delta\varphi_+$ is shown also for 
the case of a non-zero initial phase $\varphi_+^a$. (Case II)}

\end{figure}   

\end{document}